\documentclass[
  ,draft            
  ]
  {aipproc}

\layoutstyle{6x9}

\begin{document}

\title{A dynamical approach to semi-inclusive B decays}


\author{Giulia Ricciardi}{
  address={Dipartimento di Scienze Fisiche,
Universit\`a di Napoli ``Federico II'' and I.N.F.N., Sezione di
Napoli, Complesso Universitario di Monte Sant'Angelo, Via Cintia,
80126 Napoli, Italy.} }

\begin{abstract}


A dynamical approach to semi-inclusive decays based on an
effective running coupling for the strong interactions is
described.

\end{abstract}

\maketitle


\section{Introduction}

  Semi-leptonic and radiative semi-inclusive decays of heavy mesons are
currently  under intense investigation. Non perturbative physics
seems to be more manageable in such
  decays, and there is  hope for a reduction of theoretical assumptions
  and a more stringent comparison with
  experimental data.
  Several effective approaches to such decays are  available. Most commonly,
  in order to calculate the decay rate,  a series of operators is used,
  whose coefficients and matrix elements are
   weighted differently according to the
  theoretical assumptions. The matrix
  elements of the operators are not calculable within perturbation theory; the
  largest is the number of operators included in the calculation,
  the more accurate is the result.
  Another possible approach is based on an effective strong coupling, in turn based on
   perturbative threshold resumming and analyticity principles.
   \cite{Aglietti:2004fz,noi1,noi2,noi3,noi4,ugo}
   In such approach, an effective strong coupling is introduced
   and used in the resummation soft--gluon formulas.

 \section{Effective coupling}

Long distance effects manifest themselves in perturbation theory
in the form of series of large infrared logarithms, coming from
``incomplete'' cancellation of infrared divergencies in real and
virtual diagrams in the threshold kinematical region. Such
logarithms need to be resummed at all orders and resumming
formulas are available within perturbation theory. Resumming
requires integration over all possible kinematical domains,
included low energy, order of $\Lambda_{QCD}$ ones; therefore,
divergencies arise when the running coupling constant hits the
Landau pole in the integrations. Several prescriptions can be used
used in order to keep under control divergencies in the
soft--gluon resumming formulas for the decay rates. An interesting
possibility is to substitute an effective running coupling, with
no Landau pole singularity, into the resumming formulas in place
of the standard one.The non-physical Landau singularity can be
removed by means of a dispersion relation \cite{noi4}
 \begin{equation}
\label{disp_rel} \bar{\alpha}(Q^2) \, = \, \frac{1}{2\pi i}
\int_0^{\infty}  \, \frac{ds}{s \, + \, Q^2} \, {\rm Disc}_s \,
\alpha(-s);
\end{equation}
$Q$ is the hard scale of the process.
 At lowest order
  \begin{equation} \label{end}
\bar{\alpha}_{lo}(Q^2) \, = \, \frac{1}{\beta_0} \left[ \frac{ 1
}{ \log Q^2/\Lambda^2 } \, - \, \frac{ \Lambda^2 }{ Q^2 -
\Lambda^2 } \right].
\end{equation}
Let us improve the effective coupling by adding the contributions
of secondary emissions off the radiated gluons. The final
effective coupling is given by the prescription
\begin{equation}
\label{deftime2} \tilde{\alpha}(k_{\perp}^2) \, = \, \frac{i}{2
\pi} \, \int_0^{k_{\perp}^2} \, d s \, {\rm Disc}_s \, \frac{
\bar{\alpha}(- s) }{ s }.
\end{equation}
If we neglect the $-i\pi$ terms in the integral over the
discontinuity
--- i.e. the absorptive effects --- the cascade coupling exactly
reduces  to the ghost-less one:
\begin{equation}
\tilde{\alpha}(k_{\perp}^2) \, \to \, \bar{\alpha}(k_{\perp}^2).
\end{equation}

Let us notice that the dispersion relation (\ref{end}) has
automatically added a power term to the coupling. That leads to
another assumption, that is that the effective coupling may have a
role outside the perturbative contest where it has been
introduced, by including long distance effects. The perturbative
QCD formulas is extrapolated to a non-perturbative region by
assuming that the relevant non-perturbative effects can be
relegated into an effective coupling.
 By using an effective
coupling to mimic also long distance effects, one can exploit the
fact that resummation formulas have universal characteristics
which do not depend on the single process. Of course, that implies
that not all long distance effects can be accounted for. The
description is assumed valid for bound state effects; they are due
to the vibration of the $b$ quark inside the $B$ quark as a
consequence of its interactions with light degrees of freedom (the
so called Fermi motion). On the other side, other effects, like
f.i. the $K^*$ peak which appears in the radiative hadron mass
distribution  or the $\pi$ and $\rho$ peaks which appear in the
semileptonic one, cannot be accurately predicted in this approach.

\section{Comparison with data}

\begin{figure}
  \includegraphics[height=.3\textheight]{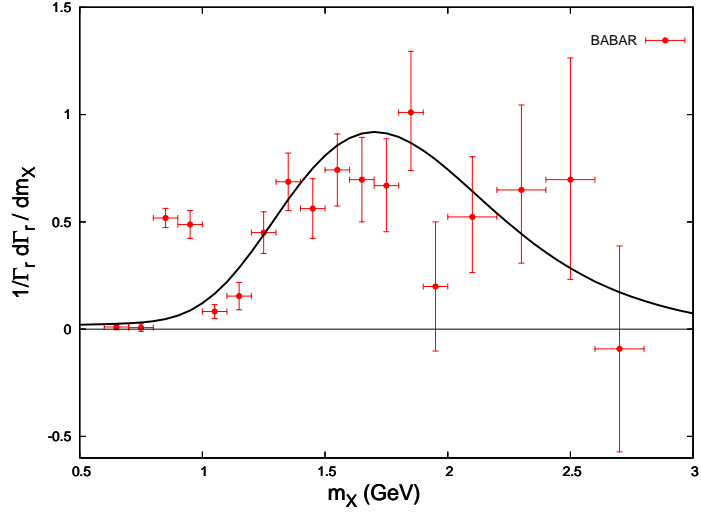}
  \caption{ \small
$B\to X_s \gamma$ invariant hadron mass distribution compared with
BaBar experimental points
 for $\alpha_S(m_Z)\, = \, 0.123$.} \label{rdmx}
\end{figure}

\begin{figure}
\includegraphics[height=.3\textheight]{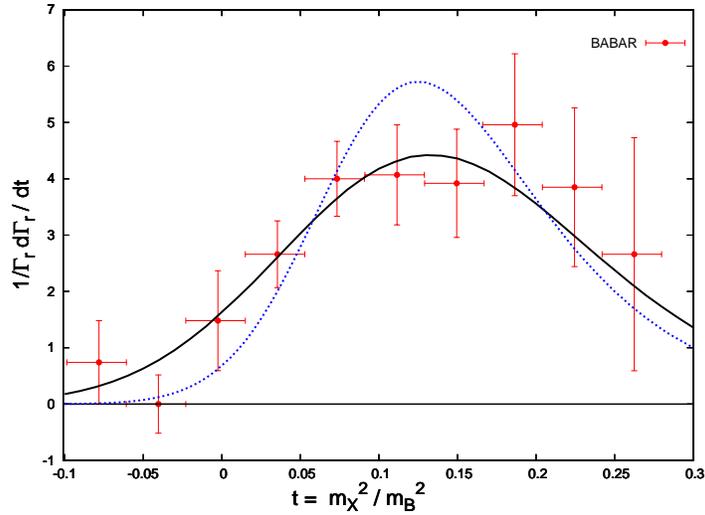} \caption{
\small $B\to X_s \gamma$ photon spectrum from BaBar compared to
the theoretical curve. The theoretical curve has been convoluted
with a normal distribution with a  standard deviation,
$\sigma_\gamma $. Here $ t \, \equiv \, 1 \, - \,
\frac{2E_{\gamma}}{m_B}$. Dotted line (blue): $\alpha_S(m_Z) \, =
\, 0.130$ and $\sigma_\gamma \, = \, 100$ MeV; continuous line
(black): $\alpha_S(m_Z) \, = \, 0.129$ and $\sigma_\gamma \, = \,
200$ MeV. } \label{efbabar}
\end{figure}

\begin{figure}
\includegraphics[height=.3\textheight]{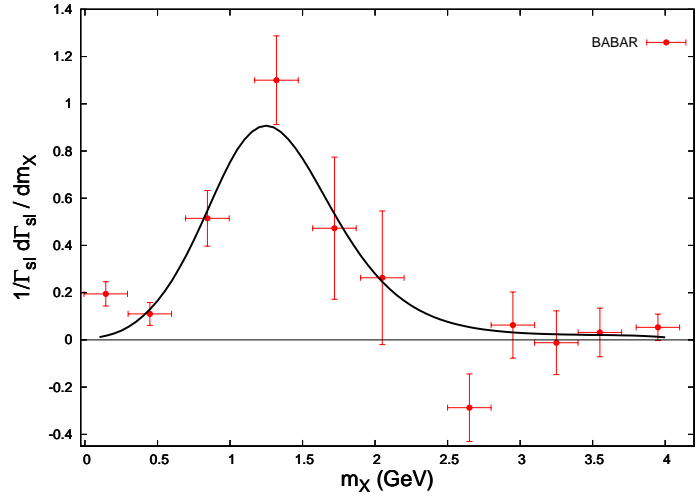} \caption{
\small invariant hadron mass distribution in semileptonic decays
from BaBar  for $\alpha_S(m_Z) \, = \, 0.119$. } \label{slmxbab}
\end{figure}

\begin{figure}
\includegraphics[height=.3\textheight]{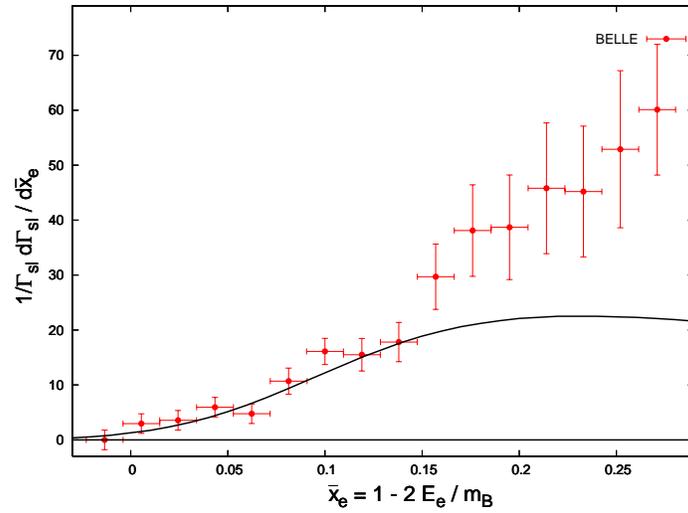} \caption{
\small electron spectrum in semileptonic decay from Belle  for
$\alpha_S(m_Z) \, = \, 0.135$.} \label{eebelle}
\end{figure}

Ultimately, the validity of the approach previously described
relies on the comparison with the experimental data. The agreement
is generally good \cite{noi4}. F.i., in Fig.~(\ref{rdmx})  and in
Fig.~(\ref{efbabar}) the invariant hadron mass distribution
$d\Gamma_r/d m_X$ and the photon energy spectrum $d\Gamma_r/d t$
for the radiative decay $B \rightarrow X_s \, \gamma$ are
compared, respectively, with experimental data given  by the BaBar
collaboration \cite{babargam2, babargam1}.
There is good agreement also for the hadron spectrum in the
semileptonic  $B \rightarrow X_u \, l \, \nu$ decay, as shown f.i.
in Fig.~(\ref{slmxbab}), where data from Babar Collaboraion are
presented \cite{Aubert:2006qi}.
The electron spectrum in the semileptonic  $B \rightarrow X_u \, l
\, \nu$ decay is affected by a large background  coming from the
decay $ B \, \to \, X_c \,  l \,  \nu $. For the electron spectrum
the agreement is not as good as in the previous cases, as can be
seen in Fig.~(\ref{eebelle}), where Belle data are shown
\cite{belleel}. However, before considering theory improvement,
one has to investigate more sophisticated comparison with data,
that are in principle possible; f.i., in this case, one could take
into account more recent data for the charm background or a better
analysis of the correlation of the systematics of the various
bins.




\bibliographystyle{aipproc}   

\hyphenation{Post-Script Sprin-ger}
\begin{thebibliography}{8}
\expandafter\ifx\csname natexlab\endcsname\relax\def\natexlab#1{#1}\fi
\providecommand{\enquote}[1]{``#1''}
\expandafter\ifx\csname url\endcsname\relax
  \def\url#1{\texttt{#1}}\fi
\expandafter\ifx\csname urlprefix\endcsname\relax\def\urlprefix{URL }\fi
\providecommand{\eprint}[2][]{\url{#2}}

\bibitem[Brown and Austin(2000{\natexlab{a}})]{Brown2000}
M.~P. Brown, and K.~Austin, \emph{The New Physique}, Publisher Name, Publisher
  City, 2000{\natexlab{a}}, pp. 212--213.

\bibitem[Brown and Austin(2000{\natexlab{b}})]{BrownAustin:2000}
M.~P. Brown, and K.~Austin, \emph{Appl. Phys. Letters} \textbf{85}, 2503--2504
  (2000{\natexlab{b}}).

\bibitem[Mittelbach and Sch{\"o}pf(1990)]{Mittelbach/Schoepf:1990}
F.~Mittelbach, and R.~Sch{\"o}pf, \emph{TUGboat} \textbf{11}, 297--305 (1990),
  \urlprefix\url{http://www.latex-project.org}.

\bibitem[Wang(2000)]{Wang}
R.~Wang, \enquote{Title of Chapter,} in \emph{Classic Physiques}, edited by
  R.~B. Hamil, Publisher Name, Publisher City, 2000, pp. 212--213.

\bibitem[van Herwijnen(1988)]{EVH:Office}
E.~van Herwijnen, {Future Office Systems Requirements}, Tech. rep., CERN DD
  Internal Note (1988).

\bibitem[Liang(1983)]{Liang:1983}
F.~M. Liang, \emph{{Word Hy-phen-a-tion by Com-pu-ter}}, Ph.D. thesis, Stanford
  University, Stanford, CA 94305 (1983), also available as Stanford University,
  Department of Computer Science Report No. STAN-CS-83-977.

\bibitem[Smith and Jones(1999)]{SJ:1999}
C.~D. Smith, and E.~F. Jones, \enquote{Load-Cycling in Cubic Press,} in
  \emph{Shock Compression of Condensed Matter-1999}, edited by M.~D.~F. et~al.,
  AIP Conference Proceedings 505, American Institute of Physics, New York,
  1999, pp. 651--654.

\bibitem[Knuth(1983)]{Knuth:WEB}
D.~E. Knuth, {The \textsf{WEB} System of Structured Documentation}, Tech. Rep.
  STAN-CS-83-980, Department of Computer Science, Stanford University,
  Stanford, CA 94305 (1983).

\end{thebibliography}


\begin{thebibliography}{9}


\bibitem{Aglietti:2004fz}
  U.~Aglietti and G.~Ricciardi,
  \emph{Phys.\ Rev.\ D} {\textbf{70}} 114008 (2004).



\bibitem{noi1}
U. Aglietti, G.~Ricciardi, and  G.~Ferrera,
  \emph{Phys.\ Rev.\ D} \textbf{74} 034004 (2006).

\bibitem{noi2}
U. Aglietti, G.~Ricciardi, and  G.~Ferrera,
  \emph{Phys.\ Rev.\ D} \textbf{74} 034005 (2006).

\bibitem{noi3}
U. Aglietti, G.~Ricciardi, and  G.~Ferrera,
  \emph{Phys.\ Rev.\ D} \textbf{74} 034006 (2006).

\bibitem{noi4}
U. Aglietti, G.~Ferrera and G.~Ricciardi,
  hep-ph/0608047.

\bibitem{ugo}
  U.~Aglietti,
  \emph{Nucl.\ Phys.\ B} \textbf{610} 293 (2001);
 U.~Aglietti, G. Corcella, G. Ferrera, hep-ph/0610035.

  \bibitem{babargam2}
  B.~Aubert {\it et al.}  [BABAR Collaboration],
  \emph{Phys.\ Rev.\ D} {\textbf{72}}  052004 (2005).
  [arXiv:hep-ex/0508004].

\bibitem{babargam1}
  B.~Aubert {\it et al.}  [BABAR Collaboration],
  arXiv:hep-ex/0507001.


\bibitem{Aubert:2006qi}
  B.~Aubert {\it et al.}  [BABAR Collaboration],
  Phys.\ Rev.\ Lett.\  {\bf 96} (2006) 221801
  [arXiv:hep-ex/0601046].

\bibitem{belleel}
  A.~Limosani {\it et al.}  [Belle Collaboration],
  Phys.\ Lett.\ B {\bf 621} (2005) 28
  [arXiv:hep-ex/0504046].


\end{thebibliography}



\end{document}